\documentclass[twocolumn,english,showpacs,showkeys]{revtex4}
\usepackage[T1]{fontenc}
\usepackage[latin9]{inputenc}
\usepackage{array}
\usepackage{graphicx}
\usepackage{setspace}

\providecommand{\tabularnewline}{\\}

\usepackage{babel}

\begin{document}

\title{Structural, static and dynamic magnetic properties of Co$_{2}$MnGe
thin films on a sapphire a-plane substrate}

\author{Mohamed Belmeguenai$^{1}$, Fatih Zighem$^{2}$, Thierry Chauveau$^{1}$, Damien Faurie$^{1}$,
Yves Roussigné$^{1}$, Salim Mourad Chérif$^{1}$, Philippe Moch$^{1}$,
Kurt Westerholt$^{3}$ and Philippe Monod$^{4}$}

\affiliation{$^{1}$ \textit{LPMTM, Institut Galilée, UPR 9001 CNRS, Université
Paris 13,} \textit{99 Avenue Jean-Baptiste Clément F-93430 Villetaneuse,
France}}

\affiliation{$^{2}$ \textit{LLB (CEA CNRS UMR 12), Centre d'études de Saclay,
91191 Gif-Sur-Yvette, France}}

\affiliation{$^{3}$ \textit{Institut für Experimentalphysik/Festkörperphysik, Ruhr-Universität Bochum, 44780 Bochum, Germany}}

\affiliation{$^{4}$ \textit{LPEM, UPR A0005 CNRS, ESPCI, 10 Rue Vauquelin, F-75231
Paris cedex 5, France}}
\begin{abstract}
Magnetic properties of Co$_{2}$MnGe thin films of different thicknesses
(13, 34, 55, 83, 100 and 200 nm), grown by RF sputtering at 400 $^{\circ}$C
on single crystal sapphire substrates, were studied using vibrating
sample magnetometry (VSM) and conventional or micro-strip line (MS)
ferromagnetic resonance (FMR). Their behavior is described assuming
a magnetic energy density showing twofold and fourfold in-plane anisotropies
with some misalignment between their principal directions. For all
the samples, the easy axis of the fourfold anisotropy is parallel
to the \textbf{c-} axis of the substrate while the direction of the
twofold anisotropy easy axis varies from sample to sample and seems
to be strongly influenced by the growth conditions. Its direction
is most probably monitored by the slight unavoidable miscut angle of
the Al$_{2}$O$_{3}$ substrate. The twofold in-plane anisotropy field
$H_{u}$ is almost temperature independent, in contrast with the fourfold
field $H_{\mathit{4}}$ which is a decreasing function of the temperature.
Finally, we study the frequency dependence of the observed line-width
of the resonant mode and we conclude to a typical Gilbert damping
constant á value of 0.0065 for the 55-nm-thick film.
\end{abstract}
\maketitle

\textbf{I.} \textbf{Introduction}\\

Ferromagnetic Heusler half metals with full spin polarization at the
Fermi level are considered as potential candidates for injecting a
spin-polarized current from a ferromagnet into a semiconductor and
for developing sensitive spintronic devices {[}1{]}. Some Heusler
alloys, like Co$_{2}$MnGe, are especially promising for these applications,
due to their high Curie temperature (905 K) {[}2{]} and to their good
lattice matching with some technologically important semiconductors
{[}3{]}. Therefore, great attention was recently paid to this class
of Heusler alloys {[}4-10{]}.

In a previous work {[}11{]}, we used conventional and micro-strip
line (MS) ferromagnetic resonance (FMR), as well as Brillouin light
scattering (BLS) to study magnetic properties of 34-nm-, 55-nm- and
83-nm-thick Co$_{2}$MnGe films at room temperature. We showed that
the in-plane anisotropy is described by the superposition of a twofold
and of a fourfold term. The easy axes of the fourfold anisotropy were
found parallel to the \textbf{c}-axis of the Al$_{2}$O$_{3}$ substrate
(and, consequently, the hard axes lie at $\pm45^{\circ}$ of \textbf{c}).
The easy axes of the twofold anisotropy were found at $\pm45^{\circ}$
of \textbf{c} for the 34-nm- and 55-nm-thick films and slightly misaligned
with this orientation in the case of the 83-nm-thick sample. However,
a detailed study of the in-plane anisotropy, involving temperature
and thickness dependence, allowing for their physical interpretation
is still missing. Therefore, it forms the aim of the present paper.
Rather complete x-rays diffraction (XRD) measurements over a large
thickness range of Co$_{2}$MnGe films are reported below in an attempt
to find correlations between in-plane anisotropies, thickness and
crystallographic textures. The thickness- and the temperature-dependence
of these anisotropies are investigated using vibrating sample magnetometry
(VSM) and the above mentioned FMR techniques. In addition, we present
intrinsic damping parameters deduced from broadband FMR data obtained
with the help of a vector network analyzer (VNA) {[}12-14{]}.\\

\textbf{I. Sample properties and preparation}\\

Co$_{2}$MnGe films with 13, 34, 55, 83, 100 and 200 nm thickness
were grown on sapphire \textbf{a}-plane substrates (showing an in-plane
\textbf{c}-axis) by RF-sputtering with a final 4 nm thick gold over
layer. A more detailed description of the sample preparation procedure
can be found elsewhere {[}11, 15{]}.

The static magnetic measurements were carried out at room temperature
using a vibrating sample magnetometer (VSM). The dynamic magnetic
properties were investigated with the help of 9.5 GHz conventional
FMR and of MS-FMR {[}11{]}. The conventional FMR set-up consists in
a bipolar \textit{X}-band Bruker ESR spectrometer equipped with a
TE$_{102}$ resonant cavity immersed is an Oxford cryostat, allowing
for exploring the 4-300 K temperature interval. The MS-FMR set-up
is home-made designed and, up to now, only works at room temperature.
The resonance fields (conventional FMR) and frequencies (MS-FMR) are
obtained from a fit assuming a Lorentzian derivative shape of the
recorded data. The experimental results are analyzed in the frame
of the model presented in {[}11{]}.\\

XRD experiments were performed using four circles diffractometers
in Bragg-Brentano geometry in order to determine $\theta-2\theta$
patterns and pole figures. The diffractometer devoted to the $\theta-2\theta$
patterns was equipped with a point detector (providing a precision
of $0.015^{\circ}$ in $2\theta$ scale). The instrument used for
recording pole figures was equipped with an Inel$^{TM}$ curved linear
detector ($120^{\circ}$ aperture with a precision of $0.015^{\circ}$
in $2\theta$ scale). The X-rays beams (Cobalt line focussource at
$\lambda=1.78897\,\AA$) were emitted by a Bruker$^{TM}$ rotating
anode. define a direct macroscopic ortho-normal reference (\textbf{1},
\textbf{2}, \textbf{3}), where the \textbf{3}axis stands for the direction
normal to the film. $\varphi$  and $\psi$ are the so-called diffraction angles
used for pole figure measurements.  $\psi$  is the declination angle between
the scattering vector and the 3-axis,  $\varphi$  is the rotational angle around the 3-axis.
The$\theta-2\theta$ patterns (not shown here) indicate that, for all the Co$_{2}$MnGe thin films, the \texttt{<}110\texttt{>}
axis can be taken along the \textbf{3}-axis. The Co$_{2}$MnGe deduced
lattice constant ($a=5.755\,\AA$ is in good agreement with the previously
published ones {[}6, 16{]}. Due to the {[}111{]} preferred orientation
of the gold over layer along the \textbf{3}-axis, only partial \{110\}
pole figures could be efficiently exploited. They behave as \{110\}
fiber textures containing well defined zones showing significantly
higher intensities (Figure 1 (a) and (b)). These regions correspond
to orientation variants which can be grouped into two families (see
Figure 1). The first one, where the threefold $[1\overline{1}1]$
or the $[1\overline{1}\overline{1}]$ axis is oriented along the \textbf{c}
rhombohedral direction, consists of two kinds of distinct domains
with the {[}001{]} axis at $\pm54.5^{\circ}$ from the \textbf{c-}axis.
The second family, which is rotated around the \textbf{3}-axis by
$90^{\circ}$ from the first one, also contains two variants. This
peculiar in-plane domain structure is presumably induced by the underlying
vanadium seed layer. As illustrated in Figure 1b, which represents
$\varphi$-scans at $\psi=60^{\circ}$, we do not observe major
differences between the crystallographic textures of the 55-nm and
of the 100-nm-thick samples : the first family shows a concentration
twice larger than the second one ; at least for the first family,
which allows for quantitative evaluations, the concentrations of the
two variants do not appreciably differ from each other; finally, about
50\% of the total scattered intensity arises from domains belonging
to these oriented parts of the scans. In the 200-nm-thick sample the
anisotropy of the fiber is less marked but the two families remain
present.\\

\begin{figure}
\includegraphics[bb=30bp 25bp 385bp 585bp,clip,width=8.5cm]{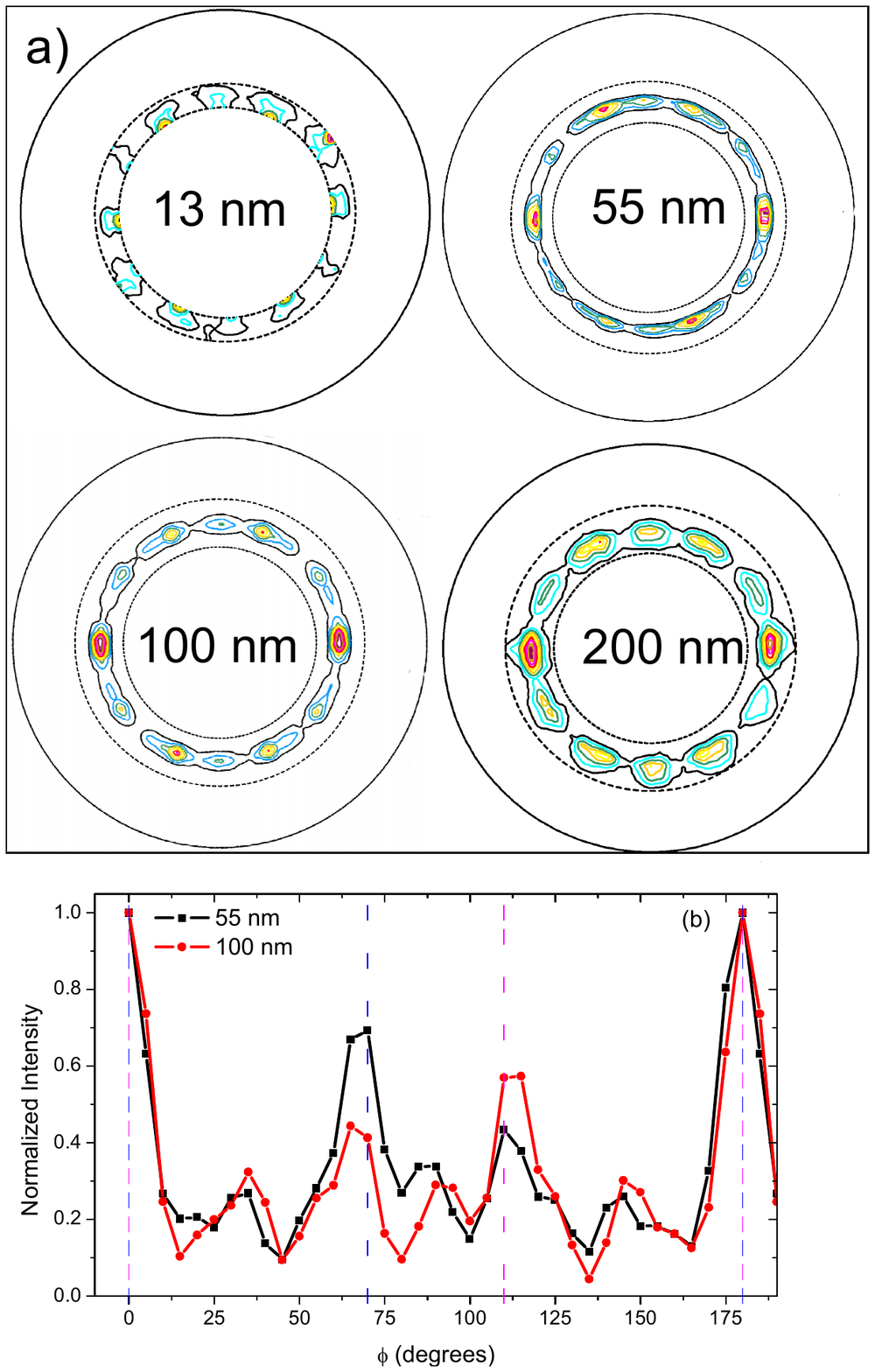}

\caption{(Color online) (a) Partial \{110\} X-rays pole figures (around 60\textdegree{})
of 13, 55, 100 and 200-nm-thick films. (b) Display of the angular
variations of the intensity derived from the above figures for the
55 and 100-nm-thick samples (the blue and pink vertical dashed lines
respectively refer to the two expected positions of the diffraction
peak relative to the two variants belonging to family 1).}

\label{Fig1:images}
\end{figure}

\textbf{III. Results and discussion}\\

\textbf{1- Static magnetic measurements}\\

In order to study the magnetic anisotropy at room temperature, the
hysteresis loops were measured for all the studied films with an in-plane
applied magnetic field along various orientations as shown in Figure2
(\textit{\ensuremath{\varphi}}$_{\mathit{H}}$ is the in-plane angle
between the magnetic applied field \textit{H} and the \textbf{c-}axis
of the substrate). The variations of the coercive field (\textit{H}$_{\mathit{c}}$)
and of the reduced remanent magnetization (\textit{M}$_{\mathit{r}}$\textit{/M}$_{\mathit{s}}$)
were then investigated as function of \textit{\ensuremath{\varphi}}$_{\mathit{H}}$.
The typical behavior is illustrated below through two representative
films which present different anisotropies.

Figure 2a shows the loops along four orientations for the 100-nm-thick
sample. One observes differences in shape of the normalized hysteresis
loops depending upon the field orientation. For \textbf{\textit{H}}
along \textbf{c}-axis ($\varphi_{H}$\textit{=0\ensuremath{^\circ}}) we
observe a typical easy axis square-shaped loop with a nearly full
normalized remanence (\textit{M}$_{\mathit{r}}$\textit{/M}$_{\mathit{s}}$=0.9),
a coercive field of about 20 Oe and a saturation field of 100 Oe.
As $\varphi_{H}$ increases away from the \textbf{c}-axis,
the coercivity increases and the hysteresis loop tends to transform
into a hard axis loop. When $\varphi_{H}$ slightly overpasses $90{}^{\circ}$
($90{}^{\circ}<\varphi_{H}<100{}^{\circ}$) the loop evolves into
a more complicated shape: it becomes composed of three (or two) open
smaller loops. Further increasing the in-plane rotation angle, it
changes from such a split-open curve up to an almost rectangular shape.
The results for $\varphi_{H}=45^{\circ}$ and $\varphi_{H}=135^{\circ}$
are different: they show a rounded loop with $M_{\mathit{r}}/M_{s}$
equal to 0.75 and 0.63 and with saturation fields of about 170 Oe
and 200 Oe, respectively. This result qualitatively agrees with a
description of the in-plane anisotropy in terms of four-fold and two-fold
contributions with slightly misaligned easy axes.\\
\begin{figure}
\includegraphics[bb=35bp 180bp 300bp 595bp,clip,width=8.5cm]{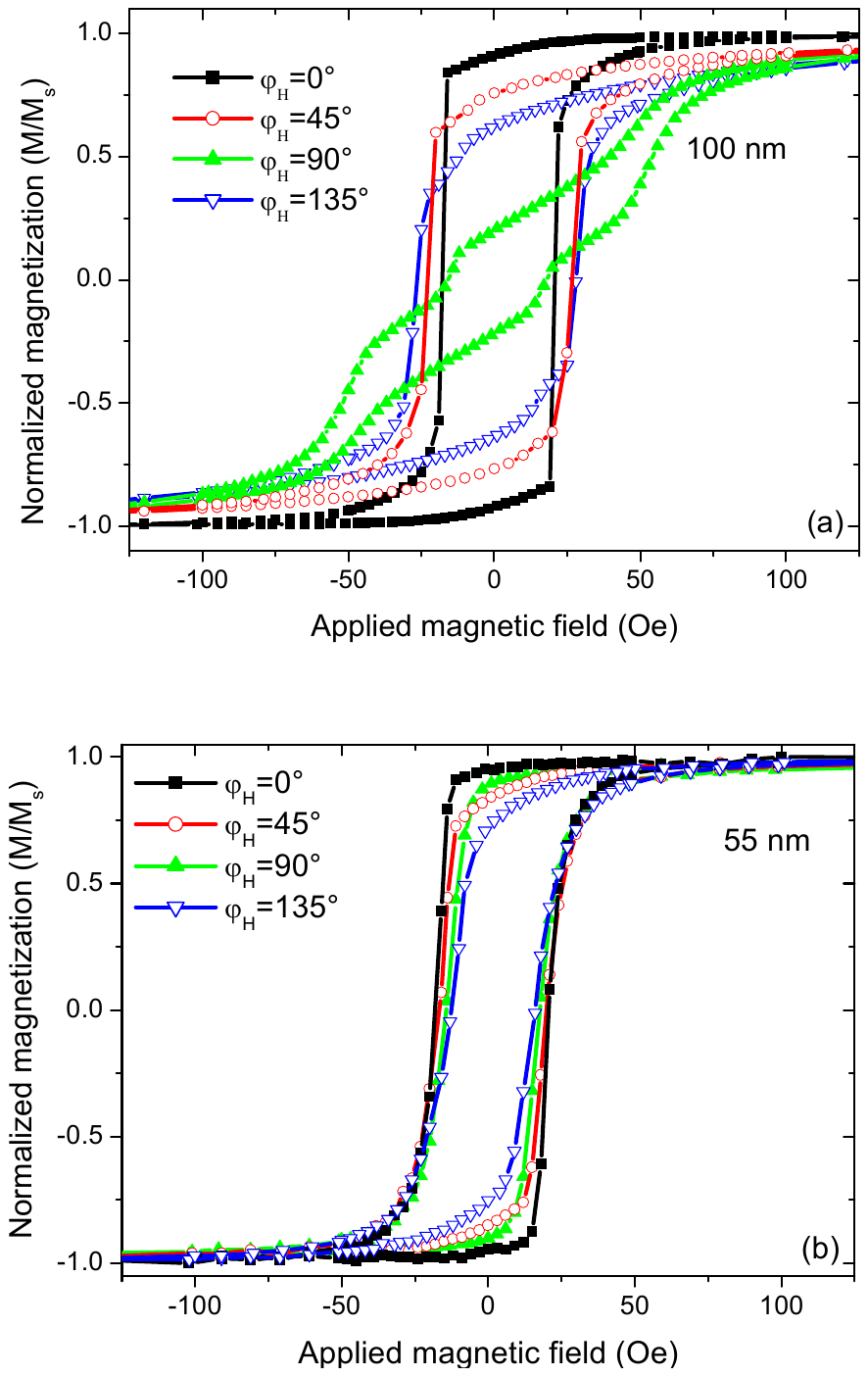}

\caption{(Color online) VSM magnetization loops of the (a) 100-nm-thick and
the (b) 55-nm-thick samples. The magnetic field is applied parallel
to the film surface, at various angles $(\varphi_{H})$ with the \textbf{c}-axis of the sapphire substrate.}

\label{Fig1:images-1}
\end{figure}

The variations of $H{}_{\mathit{c}}$ and $M_{\mathit{r}}/M_{s}$
versus $\varphi_{H}$ are illustrated in Figures 3a and 3b for the
100-nm-thick film. The presence of a fourfold anisotropy contribution
is supported by the behavior of $H_{\mathit{c}}$ (Figure 3a), since
two minima appear within each period ($180{}^{\circ}$, as expected),
as shown in Figure 3a. The minimum minimorum is mainly related to
the uniaxial anisotropy term. In the same way, as displayed in Figure
3b, the behavior of \textit{M}$_{\mathit{r}}$\textit{/M}$_{\mathit{s}}$
is dominated by the uniaxial anisotropy. It is worth to notice that
the minimum minimorum position slightly differs from $90{}^{\circ}$
(lying around $96{}^{\circ}$), thus arguing for a misalignment between
the twofold and the fourfold anisotropy axes.

Figure 2b shows a series of hysteresis loops, recorded with an in-plane
applied field, for the 55-nm-thick film. A careful examination suggests
that the fourfold anisotropy contribution is the dominant one and
that the related easy axis lies along \textbf{c}-axis. The \textit{M}$_{\mathit{r}}$\textit{/M}$_{\mathit{s}}$
variation versus $\varphi_{H}$, reported in Figure 3c, is consistent
with an easy uniaxial axis oriented at $45{}^{\circ}$ of this last
direction. Both fourfold and uniaxial terms are smaller than for the
100-nm-thick sample.\\
\begin{figure}
\includegraphics[bb=20bp 0bp 300bp 595bp,clip,width=8.5cm]{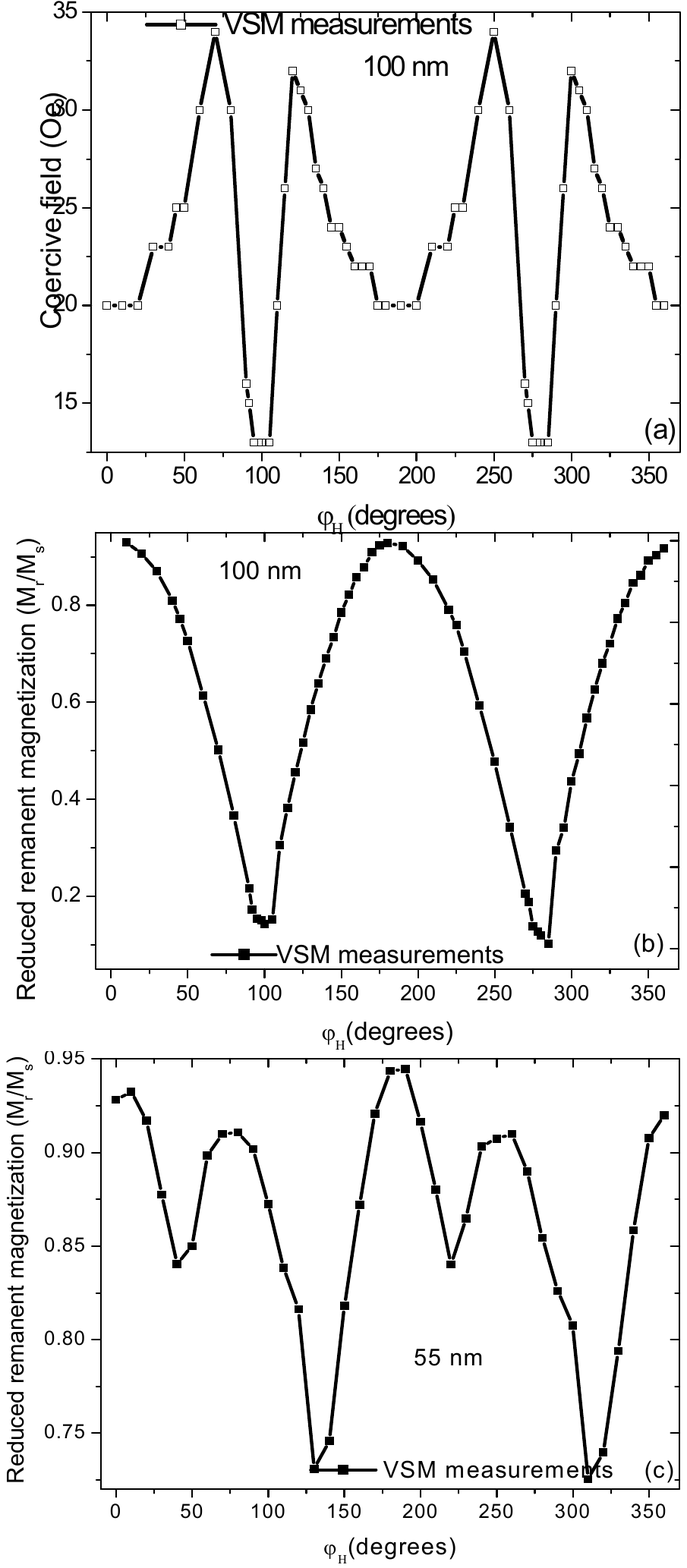}

\caption{(a) Coercive field and (b) reduced remanent magnetization of the 100-nm-thick
sample as a function of the in-plane field orientation ($\varphi_{H}$).
(c) Reduced remanent magnetization of the 55-nm-thick-film}

\label{Fig1:images-2}
\end{figure}

\textbf{2- Dynamic magnetic properties}\\

As previously published {[}11{]}, the dynamic properties are tentatively
interpreted assuming a magnetic energy density which, in addition
to Zeeman, demagnetizing and exchange terms, is characterized by the
following anisotropy contribution:\\

$E_{anis.}=K_{\perp}\sin^{2}\theta_{M}-\frac{1}{2}(1+cos(2(\varphi_{M}-\varphi_{u}))K_{u}\sin^{2}\theta_{M}-\frac{1}{8}(3+\cos4(\varphi_{M}-\varphi_{4}))K_{4}\sin^{4}\theta_{M}$
(1)\\

In the above expression, $\theta_{\mathit{M}}$ and $\varphi_{M}$
respectively represent the out-of-plane and the in-plane (referring
to the \textbf{c}-axis of the substrate) angles defining the direction
of the magnetization $M{}_{\mathbf{\mathit{s}}}$ ; $\varphi_{u}$
and $\varphi_{4}$ stand for the angles of the uniaxial axis and of
the easy fourfold axis, respectively, with this \textbf{c}-axis. With
these definitions \textit{K}$_{\mathit{u}}$ and \textit{K}$_{\mathit{4}}$
are necessarily positive. As done in ref. {[}11{]}, it is often convenient
to introduce the effective magnetization $4\pi M_{eff}=4\pi M_{s}-2K/M_{s}$,
the uniaxial in-plane anisotropy field $H{}_{\mathit{u}}=2K_{u}/M_{s}$
and the fourfold in-plane anisotropy field $H_{4}=4K{}_{\mathit{4}}/M_{s}$.\\

For an in-plane applied magnetic field \textbf{\textit{\emph{H}}},
the studied model provides the following expression for the frequencies
of the experimentally observable magnetic modes:\\
 \\
 $F_{n.}=\frac{\gamma}{2\pi}(H\cos(\varphi_{H}-\varphi_{M})+\frac{2K_{4}}{M_{s}}\cos4(\varphi_{M}-\varphi_{4})+\frac{2K_{u}}{M_{s}}\cos2(\varphi_{M}-\varphi_{u})+\frac{2A_{ex.}}{M_{s}}\left(\frac{n\pi}{d}\right)^{2})\times$\\
 $(H\cos(\varphi_{H}-\varphi_{M})+4\pi M_{eff}+\frac{K_{4}}{2M_{s}}(3+\cos4(\varphi_{M}-\varphi_{4}))+\frac{K_{u}}{M_{s}}(1+\cos2(\varphi_{M}-\varphi_{u}))+\frac{2A_{ex.}}{M_{s}}(\frac{n\pi}{d})^{2})$
(2)\\

In the above expression \textit{$\gamma$} is the gyromagnetic factor:
\textit{$(\gamma/2\pi)=g\times1.397\times10^{6}$}\textit{\emph{Hz/Oe}}.
The uniform mode corresponds to n=0. The other modes to be considered
(perpendicular standing modes) are connected to integer values of
n: their frequencies depend upon the exchange stiffness constant \textit{A}$_{\mathit{ex}}$and
upon the film thickness \textit{d}. For all the films the magnetic
parameters at room temperature were derived from MS-FMR measurements.
The deduced \textit{g} factor is equal to 2.17, as previously published
{[}11{]}.\\

The in-plane MS-FMR spectrum of the 100 nm-thick sample (Figure 4a)
submitted to a field of 520 Oe shows two distinct modes: a main one
(mode 2), with a wide line-width (about 0.6 GHz) and a second weaker
one (mode 1) at lower frequency with a narrower line-width (0.2 GHz)).
Their field-dependences are presented in Figure 4b. In contrast with
mode 2, which presents significant in-plane anisotropy, the measured
resonance frequency of mode 1 does not vary versus the in-plane angular
orientation of the applied magnetic field: such a different behavior
prevents from attributing mode 1 to a perpendicular standing excitation.
Consequently, mode 1 is presumably a uniform mode arising from the
presence of an additional magnetic phase in the film, possessing a
lower effective demagnetizing field. In the following, we focus on
mode 2 which is assumed to be the uniform mode arising from the main
phase. As previously published, only one uniform mode is observed with
the 55-nm-thick sample.\\

\begin{figure}
\includegraphics[bb=20bp 210bp 290bp 595bp,clip,width=8.5cm]{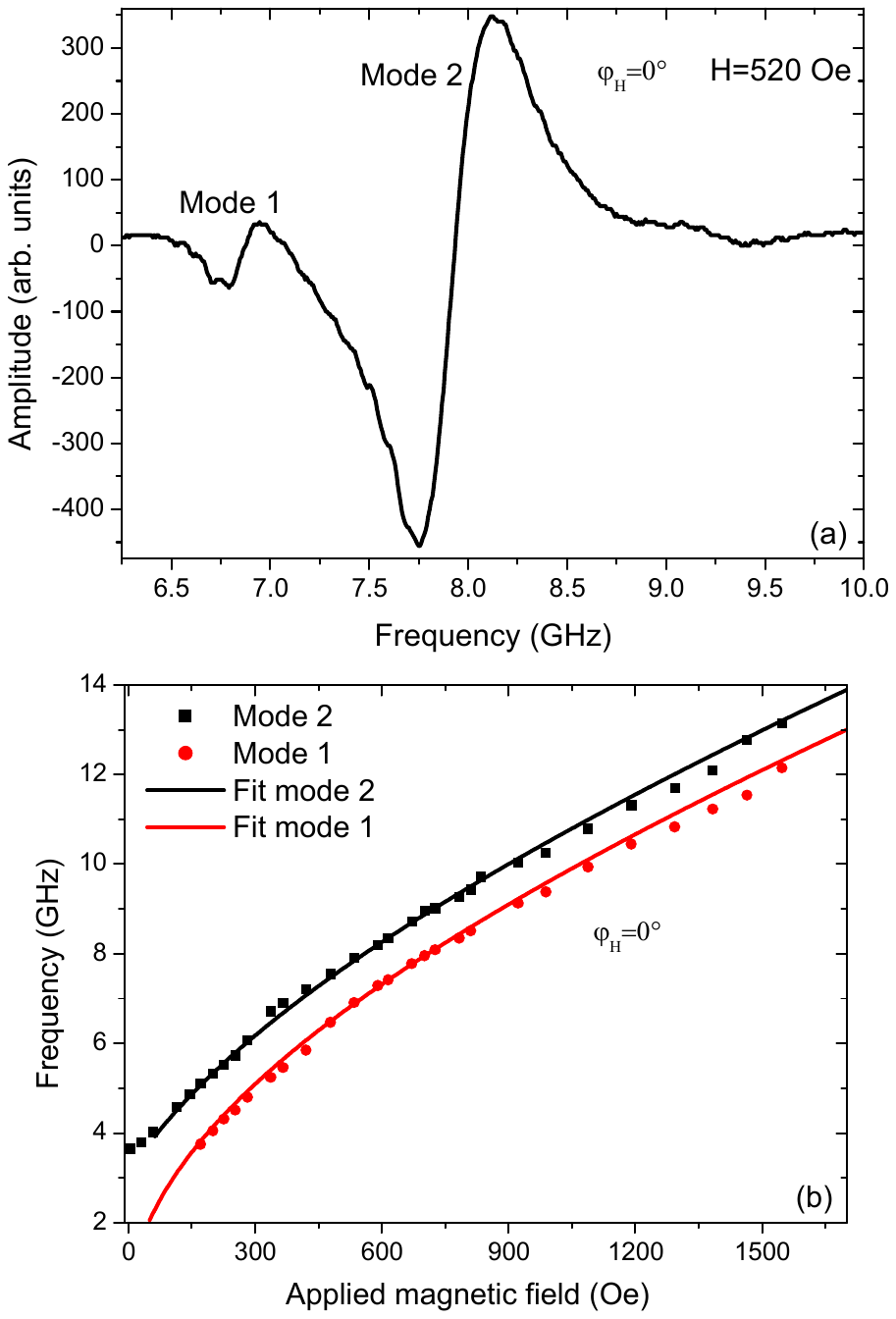}

\caption{(Color online) (a) MS-FMR spectrum under a magnetic field applied
(H=520 Oe) parallel the c-axis and (b) field-dependence of the resonance
frequency of the uniform excited modes, in the 100-nm-thick thin film.
The fits are obtained using equation (2) with the parameters indicated
in Table I. }

\label{Fig1:images-3}
\end{figure}

Figures 5b and 5d illustrate the experimental in-plane angular-dependencies
of the resonance frequency of the uniform mode for the 100- and for
the 55-nm-thick samples, compared to the obtained fits using equation
(2). As expected from the VSM measurements, in the 100-nm sample the
fourfold and uniaxial axes of anisotropy are misaligned: it results
an absence of symmetry of the representative graphs around $\varphi_{\mathit{H}}$=
$90^{\circ}$. The best fit is obtained for the following values of
the magnetic parameters: $4\pi M{}_{\mathit{eff}}=9800$ Oe, $H{}_{\mathit{u}}=55$
Oe, $H{}_{\mathit{4}}=110$ Oe, $\varphi{}_{\mathit{4}}$=0$^{\circ}$,
$\varphi_{\mathit{u}}{}^{\circ}=12{}^{\circ}$. As previously published,
in the case of the 55-nm sample the direction of the easy uniaxial
axis does not coincide with the observed one for the fourfold axis.
The best fit for this film corresponds to: $4\pi M{}_{\mathit{eff}}=9800$
Oe, $H{}_{\mathit{u}}=10$ Oe, $H{}_{\mathit{4}}=54$ Oe\textit{,}
$\varphi{}_{\mathit{4}}=0{}^{\circ}$, $\varphi_{\mathit{u}}$=45$^{\circ}$.
In both samples, the fourfold anisotropy easy direction is parallel
to the \textbf{c} axis of the substrate: this presumably results from
an averaging effect of the above described distribution of the crystallographic
orientations, in spite of the facts that such a conclusion requires
equal concentrations of the two main variants, a condition which,
strictly speaking, is not fully realized, and that the observed value
of $\varphi{}_{\mathit{4}}$does not derive from probably oversimplified
averaging model that we attempted to use, based on individual domain
contributions showing their principal axis of anisotropy along their
cubic direction.\\

As usual, attempts to interpret the in-plane hysteresis loops using
the coherent rotation model do not provide a quantitative evaluation
of the anisotropy terms involved in the expression of magnetic energy
density. However, the experimentally measured $M{}_{\mathit{r}}$\textit{/}$M{}_{\mathit{s}}$
angular variation, which, with this model, is given by $\cos(\varphi_{\mathit{M}}$\textit{-}$\varphi_{\mathit{H}})$)
in zero-applied field and is easily calculated knowing $\varphi$
$\varphi{}_{\mathit{u}}$, $\varphi_{\mathit{4}}$ and $H_{\mathit{u}}$\textit{/}$H{}_{\mathit{4}}$,
is in agreement with the values of these coefficients fitted from
resonance data, as shown in Figures 5a and 5c.\\

\begin{figure}
\includegraphics[bb=20bp 110bp 350bp 595bp,clip,width=8.5cm]{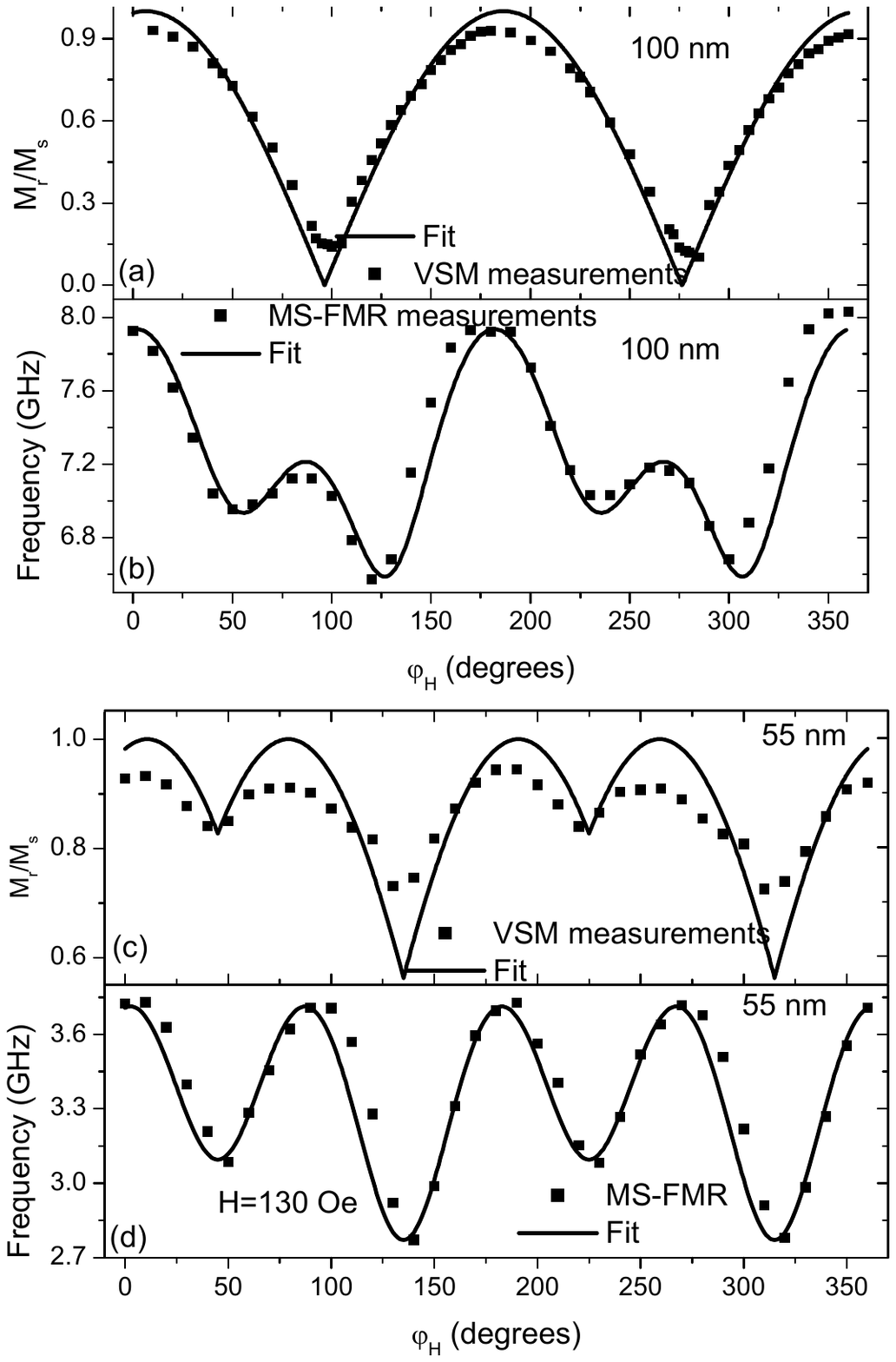}

\caption{Reduced remanent magnetization of the (a) 100-nm- and of the (c) 55-nm-thick
films. The simulations are obtained from the energy minimization using
the parameters reported in Table I. (b) and (d) show the compared
in-plane angular-dependences of the resonance frequency of the uniform
modes. The fit is obtained using equation (2) with the parameters
indicated in Table I. }

\label{Fig1:images-4}
\end{figure}

\begin{singlespace}
\noindent \begin{tabular}{>{\centering}p{1.2cm}>{\centering}p{1.5cm}>{\centering}p{1.2cm}>{\centering}m{1.24cm}>{\centering}p{1.2cm}>{\centering}p{1.2cm}}
\hline
t (nm)  & $4\pi M{}_{\mathit{eff}}$ (kG)  & $H_{u}$ (Oe)  & $H_{4}$ (Oe)  & $\varphi_{u}$ (deg.)  & $\varphi_{4}$ (deg.)\tabularnewline
\hline
\hline
13  & 8000  & 45  & 40  & 12  & 0\tabularnewline
34  & 9000  & 6  & 20  & 45  & 0\tabularnewline
55  & 9800  & 10  & 54  & 45  & 0\tabularnewline
89  & 9200  & 15  & 22  & -5  & 0\tabularnewline
100  & 9800  & 60  & 110  & 12  & 0\tabularnewline
200  & 9900  &  & 24  &  & 0\tabularnewline
\hline
\hline
\multicolumn{6}{>{\centering}p{3.3in}}{{\small Table I : Magnetic parameters obtained from the best fits
to our experimental results. $\varphi_{u}$ and $\varphi_{4}$ are
the angles of in-plane uniaxial and of fourfold anisotropy easy axes,
respectively}}\tabularnewline
 &  &  &  &  & \tabularnewline
\end{tabular}
\end{singlespace}

\noindent The magnetic parameters deduced from our resonance measurements
are given in Table I for the complete set of the studied films. In
contrast with the direction of the fourfold axis which does not vary,
the orientation of the uniaxial axis is sample dependent: for some
of them (34 and 55nm) the easy uniaxial direction lies at $45^{\circ}$
from the \textbf{c}-axis of the substrate (thus coinciding with the
hard fourfold direction); for other ones (13, 83, 100 nm) it shows
a variable misalignment; finally, the uniaxial anisotropy field vanishes for
the thickest sample (200 nm). We tentatively attribute at least a
fraction of the uniaxial contribution as originating from a slight
misorientation of the surface of the substrate. The amplitudes of
both in-plane anisotropies are sample dependent and cannot be simply
related to the film thickness. It should be mentioned that some authors
{[}17{]} have reported on strain-dependent uniaxial and fourfold anisotropies
in Co$_{2}$MnGa. This suggests a forthcoming experimental X-rays
study of the strains present in our films.\\

In addition, it is useful to get information about the damping terms
involved in the dynamics of magnetic excitations in the above samples.
Notice that in order to integrate these films in application devices
like, for instance, MRAM, it is important to make sure that their
damping constant is small enough. The damping of the 55-nm-thick film
was studied by VNA-FMR {[}12-14{]}: it is analyzed in terms of a Gilbert
coefficient $\alpha$ in the Landau-Lifschitz-Gilbert equation of
motion. The frequency line-width $\Delta f$ of the resonant signal
around $f{}_{\mathit{r}}$ observed using this technique is related
to the field line-width $\Delta H$ measured with conventional FMR
excited with a radio-frequency equal to \textit{f}$_{\mathit{r}}$
through the equation {[}18{]}:

\[
\Delta H=\Delta\frac{\partial H(f)}{\partial f}|_{f=f_{r}}(3)\]

\textit{$\Delta H$} is given by:

\[
\Delta H=\Delta H_{0}+\frac{4\pi f_{r}}{|\gamma|}\alpha(4)\]

(where \textit{$\Delta H_{0}$} stands for a small contribution arising
from inhomogeneous broadening). The measured linear dependence of
\textit{$\Delta H$} is shown versus $f_r$ in Figure 6. We then obtain the damping
coefficient: $\alpha=$0.0065. This value lies in the range observed
in the Co$_{2}$MnSi thin films {[}19-21{]}.\\

\begin{figure}
\includegraphics[bb=35bp 350bp 350bp 590bp,clip,width=8.5cm]{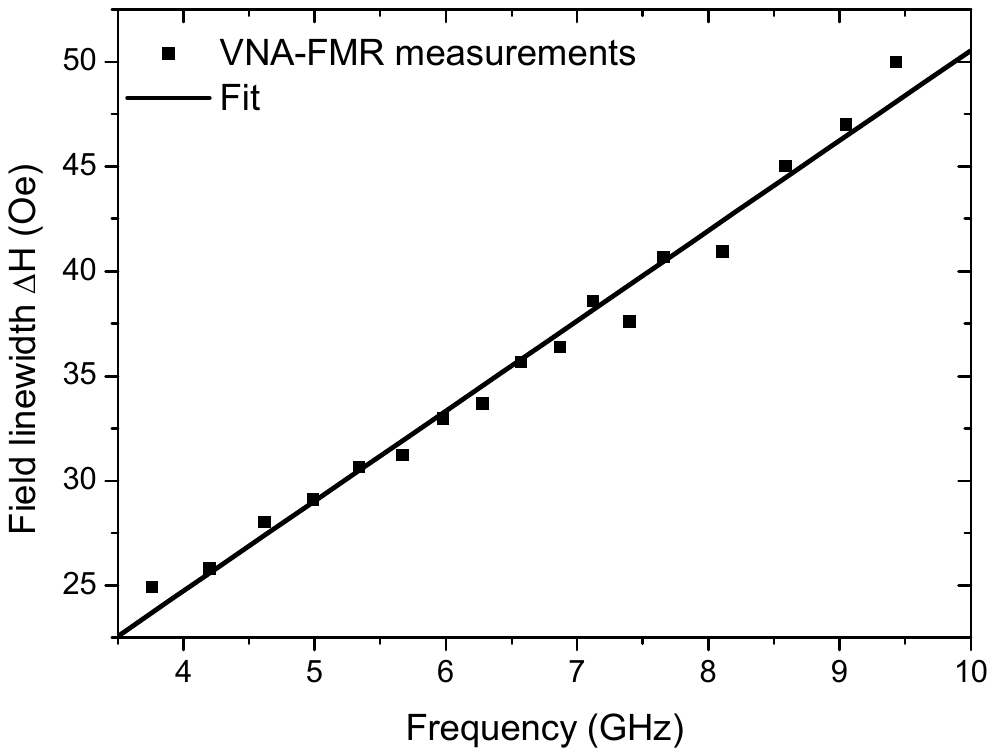}

\caption{Line-width $\Delta H$ as a function of the resonance frequency for
55-nm-thick film. $\Delta H$ is derived from the experimental VNA-FMR
frequency-swept line-width. }

\label{Fig1:images-5}
\end{figure}

Finally, the temperature dependence was studied for the 55-nm-thick
sample using conventional FMR. The fits of the magnetic parameters
were performed assuming that \textit{g} practically does not vary
versus the temperature T, as generally expected. We then take: \textit{$g=2.17$}.
The results for the uniaxial and for the fourfold in-plane anisotropy
fields are reported in Figure 7. $H_{\mathit{u}}$ is temperature
independent while $H_{\mathit{4}}$ is a significantly decreasing
function of T. This behavior of $H{}_{\mathit{4}}$ is presumably
related to the magneto-crystalline origin of this anisotropy term.\\
\begin{figure}
\includegraphics[bb=35bp 385bp 310bp 595bp,clip,width=8.5cm]{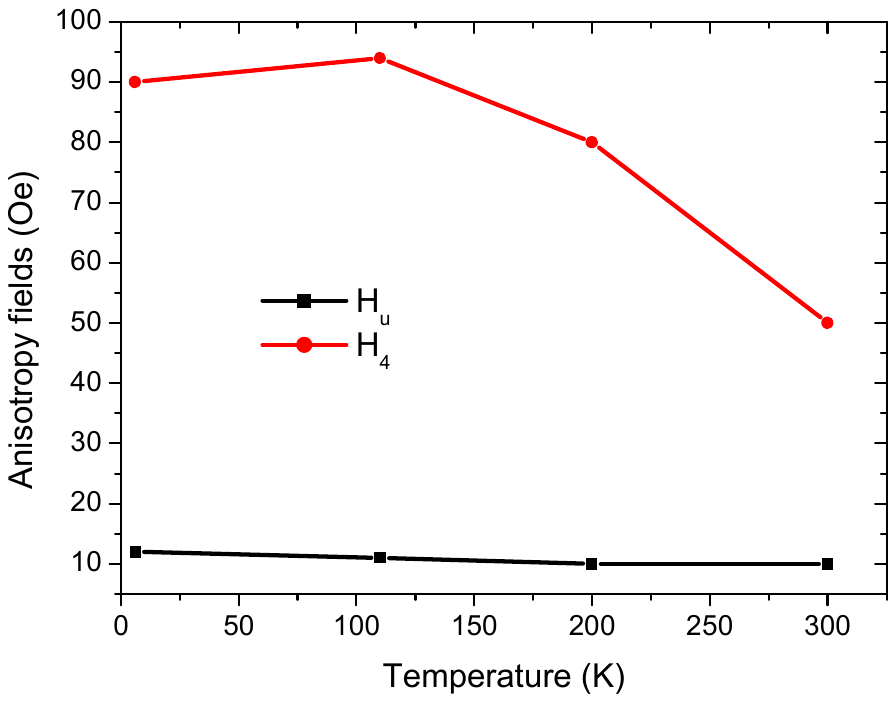}

\caption{(Color on line) Temperature-dependence of the fourfold anisotropy
field ($H_{4}$) and the unixial anisotropy field ($H_{u}$) of the
55-nm-thick film, measured by FMR at 9.5 GHz. }

\label{Fig1:images-6}
\end{figure}

\textbf{V Summary}\\

The static and dynamic magnetic properties of Co$_{2}$MnGe films
of various thicknesses sputtered on \textbf{a}-plane sapphire substrates
have been studied. The present work focused on the dependence of the
parameters describing the magnetic anisotropy upon the crystallographic
texture and upon the thickness of the films. The crystallographic
characteristics were obtained through X-ray diffraction which reveals
the presence of a majority of two distinct (110) domains. Magnetometric
measurements were performed by VSM and magnetization dynamics was
analyzed using conventional and micro-strip resonances (FMR and MS-FMR).
The main results concern the in-plane anisotropy which contributes
to the magnetic energy density through two terms: a uniaxial one and
a fourfold one. The easy axis related to the fourfold term is always
parallel to the \textbf{c-}axis of the substrate while the easy
twofold axis shows a variable misalignment with the \textbf{c}-axis. The fourfold anisotropy is a decreasing
function of the temperature: it is presumably of magneto-crystalline
nature and its orientation is related to the above noticed domains.
The observed misalignment of the two-fold axis is tentatively interpreted
as induced by random slight miscuts affecting the orientation of the
surface of the substrate. The two-fold anisotropy does not significantly
depend on the temperature. There is no evidence of a well-defined
dependence of the anisotropy versus the thickness of the films. Finally,
we show that the damping of the magnetization dynamics can be interpreted
as arising from a Gilbert term in the equation of motion, that we
evaluate.\\

\textbf{References}\\
 {\small {[}1{]} S. Tsunegi, Y. Sakuraba, M. Oogane, K. Takanashi,
Y. Ando, Appl. Phys. Lett. 93, 112506 (2008) }\\
 {\small {[}2{]} S. Picozzi, A. Continenza, and A. J. Freeman,
Phys. Rev. B 66, 094421 (2002).}\\
 {\small {[}3{]} S. Picozzi, A. Continenza, and A. J. Freeman,
J. Phys. Chem. Solids 64, 1697 (2003). }\\
 {\small {[}4{]} T. Ambrose, J. J. Krebs, and G. A. Prinz, J.
Appl. Phys. 89, 7522 (2001). }\\
 {\small {[}5{]} T. Ishikawa, T. Marukame, K. Matsuda, T. Uemura,
M. Arita, and M. Yamamoto, J. Appl. Phys. 99, 08J110 (2006)}\\
 {\small {[}6{]} F. Y. Yang, C. H. Shang, C. L. Chien, T. Ambrose,
J. J. Krebs, G. A. Prinz, V. I. Nikitenko, V. S. Gornakov, A. J. Shapiro,
and R. D. Shull, Phys. Rev. B 65, 174410 (2002).}\\
 {\small {[}7{]} H. Wang, A. Sato, K. Saito, S. Mitani, K. Takanashi,
and K. Yakushiji, Appl. Phys. Lett. 90, 142510 (2007)}\\
 {\small {[}8{]} Y. Sakuraba, M. Hattori, M. Oogane, Y. Ando,
H. Kato, A. Sakuma, T. Miyazaki, and H. Kubota, Appl. Phys. Lett.
88, 192508 (2006).}\\
 {\small {[}9{]} T. Marukame, T. Ishikawa, K. Matsuda, T. Uemura,
and M. Yamamoto, Appl. Phys. Lett. 88, 262503 (2006).}\\
 {\small {[}10{]} D. Ebke, J. Schmalhorst, N.-N. Liu, A. Thomas,
G. Reiss, and A. Hütten, Appl. Phys. Lett. 89, 162506 (2006).}\\
 {\small {[}11{]} M. Belmeguenai, F. Zighem, Y. Roussigné, S-M.
Chérif, P. Moch, K. Westerholt, G. Woltersdorf, and G. Bayreuther
Phys. Rev. B 79, 024419 (2009).}\\
 {\small {[}12{]} M. Belmeguenai, T. Martin, G. Woltersdorf, M.
Maier, and G. Bayreuther, Phys. Rev. B 76, 104414 (2007).}\\
 {\small {[}13{]} T. Martin, M. Belmeguenai, M. Maier, K. Perzlmaier,
and G. Bayreuther, J Appl. Phys. 101, 09C101 (2007)}\\
 {\small {[}14{]} M. Belmeguenai, T. Martin, G. Woltersdorf, G.
Bayreuther, V. Baltz, A. K; Suszka and B. J. Hickey, J. Phys.: Condens.
Matter 20, 345206 (2008). }\\
 {\small {[}15{]} U. Geiersbach, K.Westerholt and H. Back J. Magn.
Magn. Mater. 240, 546 (2002). }\\
 {\small {[}16{]} T. Ambrose, J. J. Krebs, and G. A. Prinz, J.
Appl. Phys. 87, 5463 (2000) }\\
 {\small {[}17{]} M. J. Pechana, C. Yua, D. Carrb, C. J. Palmstrøm,
J. Mag. Mag. Mat. 286, 340 (2005)}\\
 {\small {[}18{]} S. S. Kalarickal, P. Krivosik, M. Wu, C. E.
Patton, M. L. Schneider, P. Kabos, T. J. Silva, and J. P. Nibarger,
J. Appl. Phys. 99, 093909 (2006) }\\
 {\small {[}19{]} R. Yilgin, M. Oogane, Y. Ando, T. Miyazaki,
J. Mag. Mag. Mat. 310, 2322 (2007)}\\
 {\small {[}20{]} R. Yilgin, Y. Sakuraba, M. Oogane, S. Mizumaki,
Y. Ando, and T. Miyazaki, Japan. J. Appl. Phys. 46, L205 (2007)}\\
 {\small {[}21{]} S. Trudel, O. Gaier, J. Hamrle, and B.Hillebrands,
J. Phys. D: Appl. Phys. 43, 193001 (2010) }
\end{document}